\documentclass{article}
\usepackage{spconf,amsmath,graphicx}

\usepackage{enumitem}
\setlist{nosep, leftmargin=14pt}

\usepackage{mwe} 
\usepackage{hyperref}
\usepackage{booktabs}
\usepackage{multirow}
\usepackage[flushleft]{threeparttable}
\usepackage{array}
\usepackage{adjustbox}
\usepackage{booktabs}
\usepackage{multirow}
\usepackage{makecell}
\usepackage{rotating}
\usepackage{graphicx}
\usepackage{subcaption}
\usepackage{tabularx}
\usepackage{todonotes}
\usepackage{amsmath, amssymb}

\title{An Automated Pipeline for Tumour-Infiltrating Lymphocyte
Scoring in Breast Cancer}
\name{Adam J Shephard$^{1*}$, Mostafa Jahanifar$^{1*}$, Ruoyu Wang$^{1}$, Muhammad Dawood$^{1}$,
\textit{Simon Graham}$^{1}$, \\
\textit{Kastytis Sidlauskas}$^{2}$,
\textit{Syed Ali Khurram}$^{3}$,
\textit{Nasir M Rajpoot}$^{1}$,
\textit{Shan E Ahmed Raza}$^{1}$
}

\address{$^{1}$ Tissue Image Analytics Centre, Department of Computer Science, University of Warwick, UK \\
     $^{2}$ Barts Cancer Institute, Queen Mary University of London, UK \\
     $^{3}$ School of Clinical Dentistry, University of Sheffield, UK \\
     $^{*}$ Joint first authors contributed equally}
\begin{document}
%
\maketitle
\begin{abstract}
Tumour-infiltrating lymphocytes (TILs) are considered as a valuable prognostic markers in both triple-negative and human epidermal growth factor receptor 2 (HER2) positive breast cancer. In this study, we introduce an innovative deep learning pipeline based on the Efficient-UNet architecture to predict the TILs score for breast cancer whole-slide images (WSIs). We first segment tumour and stromal regions in order to compute a tumour bulk mask. We then detect TILs within the tumour-associated stroma, generating a TILs score by closely mirroring the pathologist's workflow. Our method exhibits state-of-the-art performance in segmenting tumour/stroma areas and TILs detection, as demonstrated by internal cross-validation on the TiGER Challenge training dataset and evaluation on the final leaderboards. Additionally, our TILs score proves competitive in predicting survival outcomes within the same challenge, underscoring the clinical relevance and potential of our automated TILs scoring pipeline as a breast cancer prognostic tool.

\end{abstract}
\section{Introduction}
\label{sec:intro}
Breast cancer is the most prevalent form of cancer worldwide, representing 15\% of new cancer cases and accounting for 7\% of cancer-related deaths in the UK \cite{cruk2022}. 
Tumour-infiltrating lymphocytes (TILs) have recently been shown to be one of the main features that plays a significant role in predicting breast cancer prognosis \cite{loi2014tumor}. 
The prognostic and predictive importance of TILs visually assessed by pathologists on biopsies and surgical resections has been shown to be significant within triple negative (TNBC) and human epidermal growth factor receptor 2 positive (HER2+) breast cancers \cite{SALGADO2015}. 
These studies have shown that an increased degree of lymphocytic infiltration is prognostic of better disease specific survival and overall survival \cite{SALGADO2015,denkert2018tils}. However, this is not the case for \textit{all} types of breast cancer.

Manual scoring of TILs exhibits significant inter/intra-rater variability, primarily due to methodological discrepancies across centers/studies, particularly in distinguishing stromal from intratumoural TILs \cite{SALGADO2015}.
The International TILs Working Group introduced a standardised approach, aiming to mitigate this variability, requiring pathologists to assess stromal TILs within the invasive tumour border alone \cite{SALGADO2015}.
Yet ambiguity persists due to challenges in precisely determining the invasive tumour boundary for TIL counting. Clearly, a more objective approach would be clinically beneficial. 

With the advances in deep learning methods over the last decade for image analysis \cite{Litjens2017, Madabhushi2016}, several methods have been proposed to segment and classify nuclei and tissue regions in Haematoxylin \& Eosin (H\&E) stained whole-slide images (WSIs) \cite{Shephard2021, AUBREVILLE2023midog, graham2023conic}. 
However, there are challenges when applying deep learning for TILs scoring. 
%
The Tumour InfiltratinG lymphocytes in breast cancER (TiGER) Challenge was launched to inspire the next generation of algorithms that can automatically and objectively generate a TILs score with high prognostic value, in HER2+ and TNBC histopathology slides. Contestants were asked to submit an algorithm to two leaderboards: Leaderboard 1 (L1) assessed the performance of the algorithm for segmenting tumour/stroma and detecting TILs in provided regions of interest (ROIs), while Leaderboard 2 (L2) assessed the prognostic capability of the generated TILs score in predicting recurrence-free survival.

\begin{figure*}[ht!]
\centering\includegraphics[width=1.0\textwidth]{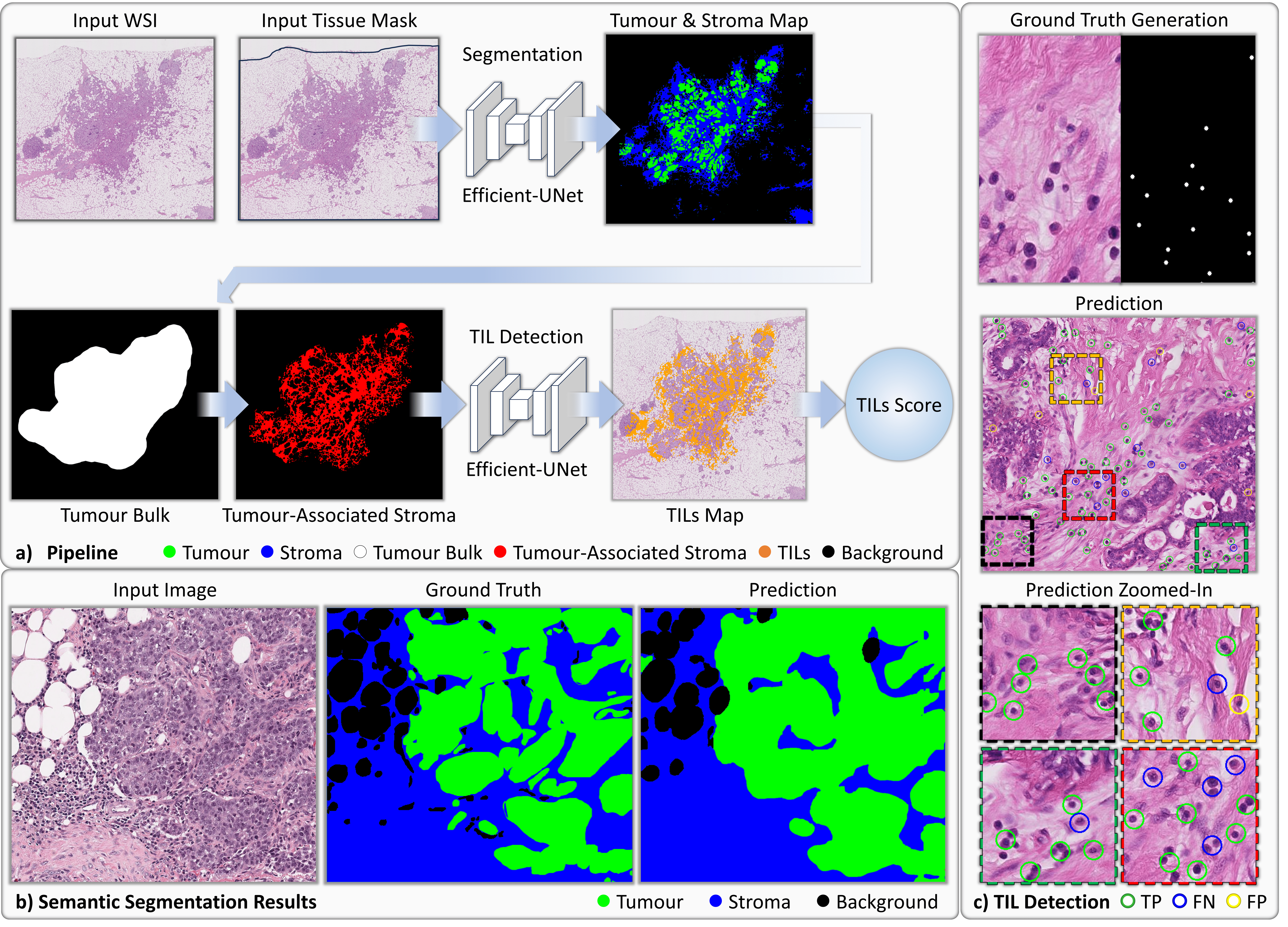}
    \caption{a) Overview of the proposed TILs Scoring pipeline. The input WSI is segmented to generate tumour-stroma segmentations. We then generate the tumour bulk to find the tumour-associated stroma. Finally, we detect the TILs in the tumour-associated stroma, and generate a WSI-level TILs score. b) Sample segmentation output with tumour, stroma and other regions. c) Sample detection output: Sample input image with ground truth dilation (top); Detections (middle) shown in circles
    ; Further zoomed-in regions (bottom).}
    \label{fig:pipeline}
\end{figure*}

In this work, we propose a fully automated pipeline  for TILs scoring based on WSIs of H\&E-stained breast cancer tissue slides. Our method is an end-to-end deep learning pipeline that has demonstrated state-of-the-art (SOTA) results in the TiGER Challenge (as team \textbf{TIAger}), particularly excelling in the segmentation of tumour/stroma and the accurate detection of TILs. We intentionally created our pipeline to closely mirror the workflow of a pathologist, ensuring interpretability and aligning with the need for human-understandable AI in medicine. To facilitate reproducibility we have made our model publicly available at \url{https://github.com/adamshephard/TIAger}. 

\section{The Proposed Method}
\label{sec:method}

We propose an end-to-end model for generating WSI-level TILs scores, using the data provided by the TiGER Challenge. The first step of our pipeline is segmenting each WSI into tumour, stroma and background. We then detect TILs in the tumour-associated stroma, where we define the tumour-associated stroma as being the stroma within the invasive tumour bulk region \cite{SALGADO2015}. Finally, we use these outputs to generate an overall TILs score for each WSI. See Figure \ref{fig:pipeline}a for an overview of the proposed pipeline.

\subsection{Study Data}

We employed the `WSIROIS' dataset provided by the challenge organisers, for training our deep learning models. This comprised of 195 WSIs of breast cancer (core-needle biopsies and surgical resections) with pre-selected ROIs and manual annotations. This dataset was curated by combining cases from three sources: \textbf{1) TCGA:} TNBC cases from TCGA-BRCA archive ($n = 151$). The annotations provided for this dataset were generated by adapting the publicly available BCSS \cite{bcss2019} and NuCLS \cite{nucls2021} datasets. \textbf{2) RUMC:} $26$ cases of TNBC and HER2+ cases from Radboud University Medical Center (Netherlands). \textbf{3) JB:} $18$ cases of TNBC and HER2+ cases from Jules Bordet Institute (Belgium). The annotations for RUMC and JB data were made by a panel of board-certified breast pathologists. All annotations included the following classes: invasive tumour, tumour-associated stroma, \textit{in-situ} tumour, healthy glands, necrosis not \textit{in-situ}, inflamed stroma and other. 

For the development and optimisation of our TILs score pipeline we used the `WSITILs' dataset, provided by the challenge organisers. This contained 82 WSIs of biopsies and surgical resections of TNBC and HER2+ breast cancer tissue from RUMC and JB. Ground truth TILs scores were provided for each WSI from a board-certified breast pathologist. All data was extracted at $20\times$ magnification ($0.5$ microns per pixel, mpp) by the organisers and provided for processing. 

As part of the TIGER Challenge, our model was tested across two leaderboards. L1 assessed the segmentation and detection quality of the model. Models were ranked according to their combined score for tumour-stroma segmentation (average of the tumour vs background and the stroma vs background Dice scores), and TILs detection (free-response receiver operating characteristic curve, FROC). L2 assessed the produced TILs score's prognostic utility. Models were ranked according to the C-index achieved using the produced TILs score in a multivariate Cox regression model trained with predefined clinical variables (age, morphology subtype, grade, molecular subtype, stage, surgery, adjuvant therapy) for predicting recurrence-free survival. On preliminary testing, L1 consisted of 26 WSIs with ROIs manually selected by the organisers, whilst L2 consisted of 200 WSIs. On final testing, L1 consisted of 38 WSIs with ROIs manually selected by the organisers, whilst L2 consisted of 707 WSIs.


\subsection{Tissue Segmentation and TILs Detection}
\subsubsection{Network Architecture}
\label{sec:architecture_combined}
To accomplish both tumour segmentation and TILs detection, we utilised the Efficient-UNet architecture, a lightweight segmentation model proposed by Jahanifar et al. \cite{jahanifar2022mitosis}, implemented in TensorFlow. The model follows an encoder-decoder design, with the encoder branch derived from the B0 variant of Efficient-Net \cite{tan2019efficientnet}, pre-trained on \textit{ImageNet}.

\subsubsection{Model Training}
\label{sec:training_combined}
We trained the proposed model to perform semantic segmentation of three tissue types: invasive tumour (`\textit{in-situ} tumour' class), stroma (`tumour-associated stroma' and `inflamed stroma' classes), and others. The training was conducted using the `WSIROIS' dataset, employing a stratified five-fold cross-validation approach. Patches of size $512\times512$ were extracted from the images at $10\times$ magnification. To enhance the model's robustness and generalisability, we employed various data augmentation techniques, including stain augmentation using the \textit{TIAToolbox} \cite{pocock2022tiatoolbox}. Initially, the decoders were trained for 10 epochs using the Adam optimizer with a learning rate of $0.003$. Subsequently, the entire network was refined for 50 epochs with a reduced learning rate ($0.0004$). For TILs detection, the same Efficient-UNet model and training strategy were utilized. The model was trained on the `WSIROIS' dataset, with an emphasis on generating binary segmentations for each TIL. This was achieved by dilating each ground truth detection by a radius of three pixels. The model was trained using patches of size $128\times128$ at $20\times$ magnification. We employed the same data augmentation and two-phase training procedure as in the segmentation task. Additionally, we implemented an on-the-fly under-sampling technique to balance the presence of patches with and without TILs in each training batch.

\begin{table}[!t]
\centering
\setlength{\tabcolsep}{5pt}
\caption{Segmentation results for Efficient-UNet on the internal cross-validation experiments.}
    \begin{threeparttable}
    \begin{tabular}{@{}cccccc@{}}
    \toprule\toprule
    \multicolumn{1}{c}{} & \multicolumn{1}{c}{} & \multicolumn{1}{c}{} & \multicolumn{3}{c}{Dice}\\
    \cline{4-6}
    \multicolumn{1}{c}{Loss} & \multicolumn{1}{c}{Patch Size} & \multicolumn{1}{c}{SN} & \multicolumn{1}{c}{Tumour} & \multicolumn{1}{c}{Stroma} & \multicolumn{1}{c}{Mean} \\ \hline
    \textbf{Jaccard} & \textbf{512} & \textbf{N} & \textbf{0.748} & \textbf{0.735} & \textbf{0.742} \\
    Jaccard & 512 & Y & 0.747 & 0.716 & 0.732 \\
    Jaccard & 1024 & N & 0.732 & 0.735 & 0.734 \\
    Dice & 512 & N & 0.734 & 0.708 & 0.721 \\
    \bottomrule\bottomrule
    \end{tabular}
\end{threeparttable}
\label{tab:segmentation-eunet}
\end{table}

\begin{table*}[!t]
\centering
\setlength{\tabcolsep}{5pt}
\caption{Segmentation and detection results of internal cross-validation experiments.}
\begin{tabularx}{0.75\textwidth}{@{}l *{6}{>{\centering\arraybackslash}X}@{}}
\toprule\toprule
\multicolumn{1}{c}{} & \multicolumn{3}{c}{Segmentation Results (Dice)} & \multicolumn{3}{c}{Detection Results} \\
\cline{2-4} \cline{5-7}
Method & Tumour & Stroma & Mean & F1 & Recall & Prec. \\
\midrule
U-Net \cite{ronneberger2015unet} & 0.627 & 0.671 & 0.649 & 0.605 & 0.780 & 0.494 \\
HoVer-Net+ \cite{Shephard2021} & 0.702 & 0.719 & 0.711 & - & - & - \\
DeepLabV3+ \cite{chen2017rethinking} & 0.703 & 0.715 & 0.723 & 0.683 & 0.755 & 0.623 \\
Swin-UNet \cite{cao2022} & 0.685 & 0.643 & 0.664 & 0.641 & \textbf{0.843} & 0.517 \\
TransUNet \cite{Chen2021} & 0.697 & 0.665 & 0.681 & 0.674 & 0.742 & 0.617 \\\midrule
\textbf{Efficient-UNet} & \textbf{0.748} & \textbf{0.735} & \textbf{0.742} & \textbf{0.702} & 0.774 & \textbf{0.642} \\
\bottomrule\bottomrule
\end{tabularx}
\label{tab:combined-results}
\end{table*}

\subsubsection{Model Inference}
The proposed model's inference was performed using an ensemble of the top-performing models from cross-validation. For tissue segmentation, patches of size $512\times512$ were extracted from tissue regions at $10\times$ magnification (256 pixels stride, 128 pixels zero-padding). The resulting segmentations were then averaged and thresholded. Morphological opening was applied to the predicted tumour region, followed by central cropping to obtain a size of $256\times256$. For TILs detection, the top three models from cross-validation were ensembled, and the output segmentations were averaged to generate a final TILs segmentation map. Tiles of size $1024\times1024$ were extracted from tissue regions at $20\times$ magnification, and these tiles were further divided into patches of size $128\times128$ ($100$ pixels stride). The model output was thresholded and individual detections were estimated via connected components.

\subsection{TILs Score}

We aimed to generate a TILs score based on the number of TILs within the tumour-associated stroma, thus mimicking the pathologist workflow. To do this, we estimated the `tumour bulk' region around the invasive tumour, using morphological operations and Delaunay triangulation, based on the tumour segmentation.
We then found the tumour-associated stroma by taking the overlap of the tumour bulk and the stroma. We performed TILs detection in this stroma, and performed WSI-level non-maxima suppression on the detections. 
Finally, we calculated the TILs score $T$,
\begin{equation}
T = \frac{NA_{\text{TILs}}}{A_{\text{TAS}}} \times 100, \quad T \in \mathbb{Z} : T \in [0, 100],
\label{eq:tilscore}
\end{equation}
where $N$ is the number of TILs within the tumour-associated stroma, $A_{\text{TILs}}$ is the area of a TIL (estimated at 16 $\mu$m), and $A_{\text{TAS}}$ is the area of the tumour-associated stroma.

\section{Results}

\subsection{Tissue Segmentation and TILs Detection}

To optimise Efficient-UNet for tissue segmentation, we tested the effect of stain normalisation (SN), patch size, and loss function (see \ref{tab:segmentation-eunet}). We found that the proposed model, using a Jaccard loss with a patch size of $512\times512$, produced the best results, giving a mean Dice of 0.742. 
We additionally compared our optimised Efficient-UNet to other SOTA models such as U-Net \cite{ronneberger2015unet},
DeepLabV3+ \cite{chen2017rethinking}, HoVer-Net+ \cite{Shephard2021, shephard2023}, Swin-UNet \cite{cao2022}, and TransUNet \cite{Chen2021} in Table \ref{tab:combined-results}. 
For detection, our model achieved an F1-score of 0.702, gaining superior results to other SOTA methods also trained based on TIL segmentations (see Table \ref{tab:combined-results}). Example segmentation and detection outputs from our model are shown in Figure \ref{fig:pipeline}b and c, respectively, showing the quality of our method.

\subsection{TILs score}
We additionally tested another SOTA model ALBRT~\cite{dawood2021albrt}, for generating a TILs score. ALBRT is a cellular composition prediction model that we adapted to predict three features (tumour area percentage, stromal area percentage, and inflammatory cell counts) from an input image patch. The mean and standard deviation of these features per slide were used as input to a Random Forest to predict the TILs score. We compared these methods based on the `WSITILs' dataset. Our method achieved a Pearson correlation coefficient, \textit{r} = 0.744, compared to \textit{r} = 0.726 by ALBRT, demonstrating the superiority of our interpretable pipeline.

\subsection{TiGER Challenge Leaderboards}
Using the proposed method in preliminary testing, we achieved a tumour-stroma mean Dice score of 0.791 and a FROC of 0.572 on L1. When tested on L2, our approach got the highest C-index of 0.719. These results obtained first place in the challenge for both the L1 and L2 preliminary leaderboards. On the final leaderboards, we gained a mean Dice score of 0.787 and a FROC of 0.544 for L1; demonstrating the robustness of our models, gaining SOTA results and second place in the competition. The winner gained a slightly increased mean Dice of 0.812 and a FROC of 0.550. For L2, all submitted methods gained substantially lower results on final testing. The C-index of our approach dropped to 0.588, with the winner of L2 gaining a C-index of 0.639. We note that the winner of L1 similarly achieved a lower performance on L2 (C-index = 0.579), showing an inverse correlation between segmentation/detection performance and survival prediction. We suggest that the substantially lower results on final testing for all submissions, may be due to differences between the preliminary and final test data, which we are unable to access.

\section{Discussion and Conclusions}

In this study, we introduced an automated and interpretable pipeline for TILs scoring in breast cancer, directly emulating the pathologist's workflow. 
Our method demonstrated SOTA performance during preliminary evaluation, securing the highest rank on the TiGER Challenge leaderboards for both the segmentation/detection and survival tasks. 
Notably, our method achieved second place for segmentation/detection on the final leaderboard. These achievements underscore the effectiveness of our pipeline in automating the complex process of TILs scoring.
Despite this, we observed a notable drop in survival prediction performance on final testing. We suspect that this discrepancy may be attributed to disparities between the preliminary and final test sets, which we did not have access to. One possible reason could be that the final dataset encompassed both biopsies and resections, which may have added further complexities. 
Furthermore, it remains unclear whether the final test set encompassed breast cancer subtypes beyond HER2+ and TNBC, where TILs are not consistently predictive.

In conclusion, our pipeline shows great promise in automating TILs scoring in a pathologist-aligned manner, addressing the crucial need for explainable AI in medical applications. Future research should focus on understanding and mitigating the variability in survival prediction results to further enhance the pipeline's robustness. The availability of our code encourages reproducibility and invites further exploration by other researchers.
\pagebreak
\clearpage

\section{Compliance with Ethical Standards}

This research study was conducted retrospectively using open access human subject data from the TiGER Challenge datasets. Ethical approval was not required.

\bibliographystyle{IEEEbib}
\bibliography{references}

\begin{thebibliography}{10}

\bibitem{cruk2022}
Cancer~Research UK,
\newblock ``Breast cancer statistics,'' https://www.cancerresearchuk.org/health-professional/cancer-statistics/statistics-by-cancer-type/breast-cancer, June 2022.

\bibitem{loi2014tumor}
Sherene Loi et~al.,
\newblock ``Tumor infiltrating lymphocytes is prognostic and predictive for trastuzumab benefit in early breast cancer: results from the finher trial.,''
\newblock {\em Annals of oncology}, 2014.

\bibitem{SALGADO2015}
R.~Salgado et~al.,
\newblock ``The evaluation of tumor-infiltrating lymphocytes (tils) in breast cancer: recommendations by an international tils working group 2014,''
\newblock {\em Annals of Oncology}, vol. 26, no. 2, pp. 259--271, 2015.

\bibitem{denkert2018tils}
Carste Denkert et~al.,
\newblock ``Tumour-infiltrating lymphocytes and prognosis in different subtypes of breast cancer: a pooled analysis of 3771 patients treated with neoadjuvant therapy,''
\newblock {\em The lancet oncology}, vol. 19, no. 1, pp. 40--50, 2018.

\bibitem{Litjens2017}
Geert Litjens et~al.,
\newblock ``{A survey on deep learning in medical image analysis},''
\newblock {\em Medical Image Analysis}, vol. 42, pp. 60--88, 2017.

\bibitem{Madabhushi2016}
Anant Madabhushi and George Lee,
\newblock ``{Image analysis and machine learning in digital pathology: Challenges and opportunities},''
\newblock {\em Medical Image Analysis}, vol. 33, pp. 170--175, 2016.

\bibitem{Shephard2021}
Adam~J. Shephard et~al.,
\newblock ``Simultaneous nuclear instance and layer segmentation in oral epithelial dysplasia,''
\newblock in {\em Proceedings of the IEEE/CVF International Conference on Computer Vision (ICCV) Workshops}, October 2021, pp. 552--561.

\bibitem{AUBREVILLE2023midog}
Marc Aubreville et~al.,
\newblock ``Mitosis domain generalization in histopathology images — the midog challenge,''
\newblock {\em Medical Image Analysis}, vol. 84, pp. 102699, 2023.

\bibitem{graham2023conic}
Simon Graham et~al.,
\newblock ``Conic challenge: Pushing the frontiers of nuclear detection, segmentation, classification and counting,''
\newblock {\em arXiv}, 2023.

\bibitem{bcss2019}
Mohamed Amgad et~al.,
\newblock ``{Structured crowdsourcing enables convolutional segmentation of histology images},''
\newblock {\em Bioinformatics}, vol. 35, no. 18, pp. 3461--3467, 02 2019.

\bibitem{nucls2021}
Mohamed Amgad et~al.,
\newblock ``Nucls: {A} scalable crowdsourcing, deep learning approach and dataset for nucleus classification, localization and segmentation,''
\newblock {\em CoRR}, vol. abs/2102.09099, 2021.

\bibitem{jahanifar2022mitosis}
Mostafa Jahanifar et~al.,
\newblock ``Stain-robust mitotic figure detection for the mitosis domain generalization challenge,''
\newblock in {\em Biomedical Image Registration, Domain Generalisation and Out-of-Distribution Analysis}, Cham, 2022, pp. 48--52.

\bibitem{tan2019efficientnet}
Mingxing Tan and Quoc Le,
\newblock ``Efficientnet: Rethinking model scaling for convolutional neural networks,''
\newblock in {\em International Conference on Machine Learning}, 2019, pp. 6105--6114.

\bibitem{pocock2022tiatoolbox}
Johnathan Pocock et~al.,
\newblock ``Tiatoolbox as an end-to-end library for advanced tissue image analytics,''
\newblock {\em Communications medicine}, vol. 2, no. 1, pp. 120, 2022.

\bibitem{ronneberger2015unet}
Olaf Ronneberger et~al.,
\newblock ``U-net: Convolutional networks for biomedical image segmentation,''
\newblock in {\em International Conference on Medical image computing and computer-assisted intervention}. Springer, 2015, pp. 234--241.

\bibitem{chen2017rethinking}
Liang-Chieh Chen, George Papandreou, Florian Schroff, and Hartwig Adam,
\newblock ``Rethinking atrous convolution for semantic image segmentation,''
\newblock {\em arXiv}, 2017.

\bibitem{cao2022}
Hu~Cao, Yueyue Wang, Joy Chen, Dongsheng Jiang, Xiaopeng Zhang, Qi~Tian, and Manning Wang,
\newblock ``Swin-unet: Unet-like pure transformer for medical image segmentation,''
\newblock in {\em European conference on computer vision}. Springer, 2022, pp. 205--218.

\bibitem{Chen2021}
Jieneng Chen et~al.,
\newblock ``{TransUNet: Transformers Make Strong Encoders for Medical Image Segmentation},''
\newblock {\em arXiv}, pp. 1--13, 2021.

\bibitem{shephard2023}
Adam~J Shephard et~al.,
\newblock ``{A Fully Automated and Explainable Algorithm for the Prediction of Malignant Transformation in Oral Epithelial Dysplasia},''
\newblock {\em arXiv}, pp. 1--35, 2023.

\bibitem{dawood2021albrt}
Muhammad Dawood et~al.,
\newblock ``Albrt: Cellular composition prediction in routine histology images,''
\newblock in {\em Proceedings of the IEEE/CVF ICCV}, 2021, pp. 664--673.

\end{thebibliography}

\end{document}